\def\papermode{2}
\ifnum\papermode=1
    \documentclass[twocolumn,twocolappendix,trackchanges]{aastex7}
\else 
    \documentclass[twocolumn,twocolappendix]{aastex7}
\fi

\usepackage{amsmath,amstext}
\usepackage{CJK}
\usepackage[T1]{fontenc}
\usepackage{newtxtext,newtxmath}
\usepackage[figure,figure*]{hypcap}
\usepackage{booktabs}
\usepackage{cleveref}
\usepackage{paralist}
\usepackage{ulem}

\newcommand{\refsec}[1]{\S\ref{#1}}

\usepackage{soul}

\begin{document}
\begin{CJK*}{UTF8}{gbsn}

\title{Evolution of the stellar mass function in open clusters from a universal and unsegregated initial state}

\author[orcid=0000-0002-0880-3380]{Lu Li (李璐)}
\affiliation{Shanghai Astronomical Observatory, Chinese Academy of Sciences, 80 Nandan Road, Shanghai 200030, China.}
\email{lilu@shao.ac.cn}  
\altaffiliation{Corresponding authors: Lu Li (\href{mailto:lilu@shao.ac.cn}{lilu@shao.ac.cn}) and Zhaozhou Li (\href{mailto:zhaozhou.li@nju.edu.cn}{zhaozhou.li@nju.edu.cn})} 

\author[0000-0001-7890-4964]{Zhaozhou Li (李昭洲)}
\affiliation{School of Astronomy and Space Science, Nanjing University, Nanjing, Jiangsu 210093, China}
\affiliation{Key Laboratory of Modern Astronomy and Astrophysics, Nanjing University, Ministry of Education, Nanjing 210093, China}
\affiliation{Centre for Astrophysics and Planetary Science, Racah Institute of Physics, The Hebrew University, Jerusalem, 91904, Israel}
\email{zhaozhou.li@nju.edu.cn}

\author[0000-0001-8611-2465]{Zhengyi Shao (邵正义)}
\affiliation{Shanghai Astronomical Observatory, Chinese Academy of Sciences, 80 Nandan Road, Shanghai 200030, China.}
\affiliation{Key Lab for Astrophysics, Shanghai 200234, China}
\email{zyshao@shao.ac.cn}

\author[0009-0007-2280-1254]{Zepeng Zheng (郑泽鹏)}
\affiliation{School of Physics and Astronomy, Sun Yat-sen University, Daxue Road, Zhuhai 519082, China}
\email{zhengzp3@mail2.sysu.edu.cn}

\author[0000-0001-8713-0366]{Long Wang (王龙)}
\affiliation{School of Physics and Astronomy, Sun Yat-sen University, Daxue Road, Zhuhai 519082, China}
\affiliation{CSST Science Center for the Guangdong-Hong Kong-Macau Greater Bay Area, Zhuhai 519082, China}
\email{wanglong8@mail.sysu.edu.cn}


\begin{abstract}

The stellar mass function (MF) and its spatial variation (mass segregation) within star clusters encode signatures of early formation physics and subsequent secular evolution. Yet, a coherent evolutionary picture remains elusive due to conflicting reports regarding the universality of the initial mass function (IMF) and the prevalence of primordial mass segregation. These discrepancies often arise from unresolved binaries, field contamination, and completeness bias. Here, we resolve these issues by analyzing 163 high-fidelity open clusters via a Bayesian forward-modeling framework. We reveal a remarkably simple initial state: young clusters ($\lesssim 300$ Myr) share a mean IMF slope of $-2.29$ in the mass range $M \geq 0.5 M_\odot$, consistent with the Salpeter slope but with an intrinsic scatter of 0.17, and exhibit minimal mass segregation at the onset of gas-free evolution ($\sim$10 Myr). This broadly universal "zero-point" for secular evolution disfavors star-forming scenarios that predict strong primordial segregation or significant IMF variations, and suggests that chaotic cluster assembly and gas expulsion efficiently erase any mild primordial inhomogeneities. By tracing the evolutionary sequence from $10^7$ to $10^{9.8}$ yr, we demonstrate that dynamical processing operates on distinct timescales: mass segregation proceeds rapidly via internal relaxation, whereas global MF flattening due to tidal evaporation becomes dominant only after $\sim$600 Myr. These findings impose robust observational constraints on the physics of star formation and early feedback and establish an empirical baseline for modeling secular stellar dynamics.
\end{abstract}

\section{Introduction}
\label{sec:intro}

Star clusters are fundamental building blocks of galaxies and essential laboratories for testing the physics of star formation and feedback.
Their present-day stellar mass function and its spatial variation result from the interplay of early formation \citep{Kroupa2001,Lada2003,Hennebelle2024} and subsequent secular evolution such as internal relaxation and external tides.
Specifically, competing paradigms predict distinct correlations between the initial mass function (IMF) and primordial mass segregation \citep{Salpeter1955,Kroupa2001,Bastian2010}.
Competitive accretion concentrates growth in deep potential wells, implying strong primordial segregation and allowing for environmental IMF variations \citep{Bonnell1998,Bastian2010}.
Conversely, turbulent fragmentation and monolithic collapse favor a stochastic, unsegregated initial configuration \citep{Parker2014}, with an IMF that is either universal \citep{Padoan2002} or sensitive to initial conditions \citep{Girichidis2012}.
In modern global hierarchical collapse scenarios, both signatures are also shaped by the assembly history of sub-clusters \citep{Vazquez-Semadeni2019,Polak2025}.
These primordial features are further modulated by early dynamical relaxation and gas expulsion \citep{Baumgardt2007,McMillan2007,Allison2009,Dominguez2017,Pavlik2019b}.
Therefore, establishing an observational ``zero-point'' at the onset of gas-free evolution discriminates between these scenarios, while tracing the subsequent evolution of older clusters calibrates the dynamical clock and disentangles secular dynamics operating on distinct timescales \citep{Baumgardt2003,Webb2016}.

However, constructing this evolutionary chronology remains observationally elusive. Young clusters directly probe early states. While broadly consistent with the canonical Salpeter value ($\alpha=-2.35$ at high-mass end), their reported mass-function slopes exhibit substantial scatter (e.g., $\alpha \sim -3.7$ to $-1.7$ for clusters $\lesssim 100\,$Myr, \citealt{Dib2014, Ebrahimi2022, Alfonso2024}), which could reflect either physical variations or measurement systematics.
Observations of some protoclusters and young clusters reveal signatures of mass segregation \citep{2024Xu, Alfonso2024}, yet whether this reflects a methodological artifact \citep[e.g.\ of minimum-spanning-tree estimators;][]{Parker2015}, a primordial imprint \citep{Pavlik2019}, or rapid dynamical evolution \citep{Allison2009} remains debated. In older clusters, tidal evaporation of low-mass stars and two-body relaxation hinder the backward inference of initial conditions, making forward $N$-body simulations an essential complement \citep[e.g.,][]{Baumgardt2008}.

A major concern in interpreting these measurements arises from a fundamental trade-off between completeness and purity in member selection \citep{Bastian2010}. Methods prioritizing kinematic purity inadvertently exclude high-velocity low-mass stars in cluster outskirts, artificially flattening the mass function \citep{Angelo2023}. Conversely, relaxed criteria introduce field-star contamination that steepens the slope \citep{Sollima2017b}, while unresolved binaries cause further flattening when misinterpreted as higher-mass stars \citep{MaizApellaniz2008, Li2020}. Moreover, many studies probe different mass ranges, complicating direct quantitative comparisons \citep{Bastian2010}. Resolving these ambiguities requires a framework that simultaneously accounts for these competing biases, applied to a large homogeneous sample, over a broad span of ages and within a common stellar mass range.

In this work, we address these challenges by analyzing the ``MF Prime'' sample, a high-fidelity subset of 163 clusters from the MiMO catalog \citep{Li2025}. 
Utilizing Gaia DR3 and the Bayesian forward-modeling framework MiMO \citep{Li2022h}, we robustly infer the cluster age, mass function ($\alpha$ for $M \ge 0.5\,M_\odot$), and its spatial variation ($\Delta \alpha$).
Unlike traditional approaches, our method treats the color-magnitude diagram (CMD) as a probabilistic mixture of single stars, unresolved binaries, and field contaminants, thereby avoiding rigid membership cuts and recovering members across the entire cluster extent without compromising purity. 

Our analysis reveals that clusters in the local Galactic disk emerge from the natal gas phase with a near-canonical Salpeter slope and an effectively unsegregated structure. 
We demonstrate that the strong segregation and flattened mass functions observed in older systems are predominantly evolutionary consequences, driven by rapid internal relaxation and delayed tidal evaporation ($\gtrsim$600\,Myr), a scenario further validated through direct $N$-body simulations.

\section{Methodology}
\label{sec:method}

\subsection{Data sample and selection}
\label{sec:data_select}
Our analysis relies on the ``MF Prime'' sample, a high-fidelity subset of 163 open clusters selected from the comprehensive MiMO catalog \citep{Li2025} for their exceptional data quality. 
The parent catalog comprises 1,232 clusters with physical parameters derived from Gaia DR3 astrometry and photometry \citep{Riello2021} using the Mixture Model for Open clusters (MiMO) framework \citep{Li2022h}.
The data and code are publicly available at the China-VO PaperData service \citep{MiMOcat}.
To ensure high completeness for member stars even in the cluster outskirts, the input data for each cluster include all stars within a broad spatial and proper motion window (typically $3 \times$ the projected half-number radius and $6 \times$ the velocity dispersion) \citep{Li2025}.

From the comprehensive MiMO catalog, the prime sample was selected through inspection of Gaia color-magnitude diagrams (CMDs), prioritizing clusters with narrow, well-defined main sequences and clear separation from the neighboring field population to ensure the most reliable stellar mass function (MF) estimates. 
The resulting sample spans an age range of $10^7$ to $10^{9.8}$\,yr and exhibits an unbiased age distribution representative of the full catalog (as explicitly shown in Figure 2 of \citealt{Li2025}; see \citealt{Hunt2026} for general discussion on completeness).
While the selection criteria favor nearer clusters with low extinction, this wide age coverage makes the sample representative for quantifying cluster evolution in the local Galactic environment.

\subsection{Bayesian inference with MiMO}
\label{sec:mimo}

We employ the Bayesian forward-modeling framework MiMO to infer physical parameters directly from the observed CMD \citep[see][for details]{Li2022h, Li2025}.
Unlike traditional methods that rely on pre-selecting members via strict kinematic cuts (which may lead to ``completeness bias'' by disproportionately removing low-mass stars with high velocity dispersions), MiMO models the observed data distribution in the CMD (magnitude $m$, color $c$) as a probabilistic mixture of two populations:
\begin{equation}
    p(m, c | \boldsymbol{\Theta}) = (1 - f_{\text{fs}}) \, \phi_{\text{cl}}(m, c | \boldsymbol{\Theta}) + f_{\text{fs}} \, \phi_{\text{fs}}(m, c),
\end{equation}
where $f_{\text{fs}}$ is the field-star fraction, and $\phi_{\text{fs}}$ is the empirical field density constructed from neighboring sky regions.
The cluster model, $\phi_{\text{cl}}$, is determined by the theoretical isochrone (defined by age, metallicity, distance, and extinction), the stellar mass function, and the binary properties (fraction and mass ratio distribution) convolved with observational errors.
The model vector $\boldsymbol{\Theta}$ encapsulates all these cluster parameters.
Specifically, we adopt PARSEC 1.2S isochrones \citep{Bressan2012b} with the Gaia EDR3 photometric system \citep{Riello2021} and the YBC bolometric corrections \citep{Chen2019c}.
Using nested sampling \citep{Skilling2004a,Speagle2020}, MiMO robustly retrieves the cluster parameters $\boldsymbol{\Theta}$, including the age and stellar mass function slope.

Crucially, by explicitly modeling field stars and unresolved binaries as mixture components, we avoid the contamination biases that commonly affect MF measurements. 
Unaccounted-for binaries artificially flatten the inferred MF by mimicking higher-mass stars, whereas field contamination steepens it through the abundance of faint background sources \citep{Li2022h}. 
This probabilistic approach is also advantageous for age determination: by retaining more stars near the main-sequence turn-off that traditional kinematic cuts might inadvertently exclude (particularly in older distant clusters), MiMO yields age estimates with demonstrated high precision and robustness \citep{Li2025}.
These improvements in MF and age measurements are critical for resolving the subtle evolutionary trends presented in this work.

\subsection{Measurement of mass function and segregation}
\label{sec:mf_seg}

To ensure consistency across the sample, we model the stellar mass function within the MiMO framework as a power law ($dN/dM \propto M^{\alpha}$) for stars with $M \geq 0.5~M_{\odot}$ and measure the slope $\alpha$.
This mass regime captures the bulk of the stellar mass, where both the canonical initial mass function (IMF) \citep{Salpeter1955,Kroupa2001} and the observed present-day MF of clusters across a wide range of ages \citep{Li2022h} are well described by a single power law.
Here the MF refers to the masses of single stars and the primaries of unresolved binaries; their secondaries are accounted for separately through the binary fraction $f_\mathrm{b}$ and mass-ratio distribution, both fitted simultaneously within MiMO.

We implement this lower mass threshold via a magnitude cut derived from the best-fit isochrone of each cluster \citep{Li2025}. 
We impose no explicit upper mass limit, allowing stellar evolution to naturally truncate the high-mass end.

To quantify mass segregation, we partition each cluster into inner and outer regions based on the projected half-number radius, $r_{50}$, derived from the MiMO membership probabilities.
We apply the MiMO inference independently to each region to determine the inner ($\alpha_{\text{in}}$) and outer ($\alpha_{\text{out}}$) slopes.
The degree of mass segregation is defined as $\Delta \alpha = \alpha_{\text{in}} - \alpha_{\text{out}}$.

\begin{figure*}[t]
    \centering
    \includegraphics[width=0.85\textwidth]{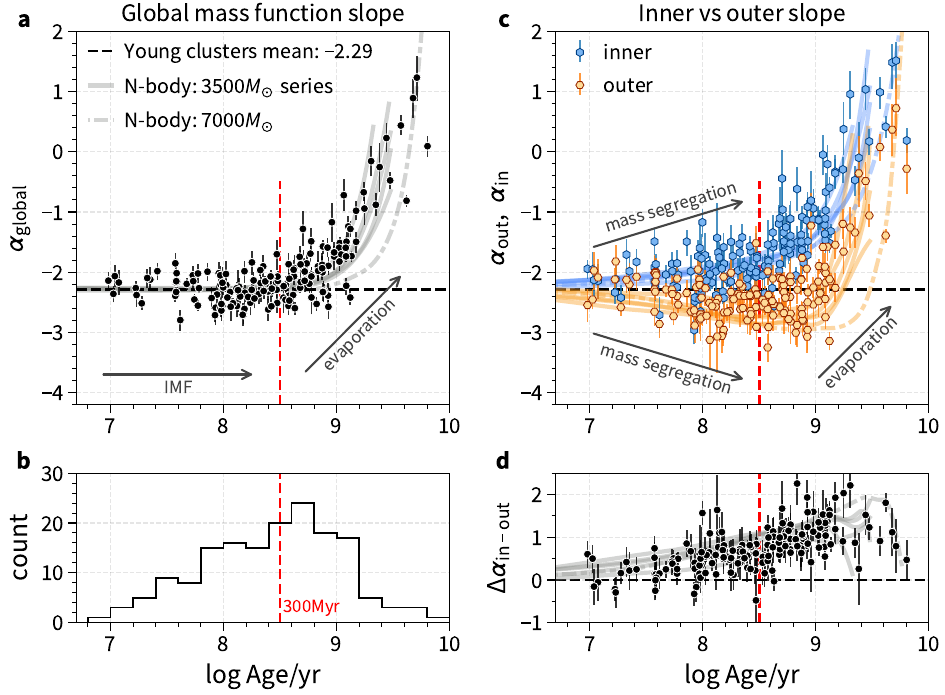}
    \vspace{-0.3em}
    \caption{%
    Evolution of the stellar mass function of observed open clusters.
    \textbf{(a):} The global mass function slope ($\alpha_{\text{global}}$) as a function of age for 163 open clusters (black points).
    Young clusters ($< 300$ Myr) show a tight distribution around the mean $\alpha = -2.29$ (black dashed line) with an intrinsic scatter of $0.17$ (after correcting for measurement uncertainties), indicating a nearly universal initial condition consistent with the canonical Salpeter slope $\alpha = -2.35$.
    A distinct upturn (flattening) due to tidal evaporation (preferential loss of low-mass stars) becomes evident only after $\sim 600$ Myr. 
    \textbf{(b):} The age distribution of the sample, covering the full evolutionary range from the onset of the gas-free phase to dissolution.
    \textbf{(c):} Differential evolution of the mass function slope in the inner (blue, inside the half-number radius $r_{50}$) and outer (orange) regions.
    \textbf{(d):} The degree of mass segregation, $\Delta \alpha = \alpha_{\text{in}} - \alpha_{\text{out}}$. 
    Mass segregation is minimal at an age of $\sim 10$ Myr but increases steadily thereafter, driven by rapid internal relaxation.
    Overlaid in panels (a, c, d) are evolutionary tracks from direct $N$-body simulations initialized with a Kroupa IMF and no primordial mass segregation (solid pale curves, color-coded to match data: initial mass of $3500 M_{\odot}$; dot-dashed: $7000 M_{\odot}$), which reproduce the observed timescales for segregation and evaporation.
    }
    \vspace{1em}
    \label{fig:mf}
\end{figure*}

\section{Results}
\label{sec:results}

\subsection{Young clusters: Broadly universal and unsegregated initial state}
\label{sec:init_state}
By focusing on the youngest clusters in our sample ($< 300$ Myr), where dynamical evaporation is minimal, we constrain the initial conditions of cluster formation.
In the mass regime $M > 0.5 M_{\odot}$, the global mass function slopes distribute tightly around a mean value of $\alpha_{\text{global}} = -2.29$ with an intrinsic scatter of $\sigma_{\text{int}} = 0.17$ (after accounting for individual measurement errors with an RMS of $\sigma_{\text{err,RMS}} = 0.16$; Fig.~\ref{fig:mf}a).
This mean value is statistically consistent with the canonical Salpeter slope ($\alpha = -2.35$) and the Kroupa IMF \citep{Salpeter1955, Kroupa2001}.
While the intrinsic scatter warrants future investigation into second-order environmental or stochastic effects (e.g., variations in local density or turbulence), the robustness of the mean suggests that the primary physics governing star formation is largely insensitive to local variations within the Galactic disk, and that the violent gas expulsion phase does not preferentially remove stars of certain masses.

Moreover, at the onset of gas-free evolution ($\sim 10$ Myr, \citealt{Krause2016}), we detect only minimal to mild mass segregation, with a relatively small difference between inner and outer MF slopes ($\Delta \alpha = \alpha_{\text{in}} - \alpha_{\text{out}}$) (Fig.~\ref{fig:mf}c).
Although slight segregation signatures ($\Delta \alpha\simeq 0.5$) appear in some clusters, they are quantitatively consistent with rapid early relaxation acting on an initially unsegregated population (given the measurement uncertainties), as confirmed by our $N$-body simulations (see below), rather than the strong primordial segregation predicted by competitive accretion models \citep{Bonnell1998}.
Instead, our results support scenarios where stars form with a homogeneous spatial distribution or where early substructure is efficiently erased during the gas-embedded phase \citep{Allison2009,Pavlik2019b}.

This unsegregated configuration also explains the preservation of the Salpeter slope reported above: had the clusters been primordially segregated with low-mass stars in the outskirts, the violent gas expulsion phase would have preferentially removed them, thereby flattening the global MF \citep{2008Marks,Baumgardt2008}.
Therefore, ${\sim}10$ Myr establishes an empirical zero-point of largely preserved IMF and weak segregation for secular dynamical evolution.

\subsection{Older clusters: Segregation vs tidal stripping}
\label{sec:seg_evap}

Starting from the baseline of young clusters, we resolve the dynamical history of open clusters by decoupling the timescales of two competing processes.
Driven by two-body relaxation, internal mass segregation develops on a timescale shorter than evaporation, causing low-mass stars to migrate outward and initially steepen $\alpha_{\text{out}}$ (Fig.~\ref{fig:mf}b).
During this early phase, the global slope $\alpha_{\text{global}}$ remains robustly Salpeter, as stars are merely redistributed internally rather than lost.
However, a critical transition occurs at $\sim 600$ Myr when tidal stripping begins to efficiently deplete these loosely bound low-mass stars.
This evaporation drives the reversal of the steepening trend in the outskirts and \textit{simultaneously} triggers the flattening of the global mass function \citep{Baumgardt2003} (as characterized by piecewise linear fitting; see below).
These synchronized features confirms the theoretical expection that the timescale for segregation is significantly shorter than for dissolution (see Appendix \ref{sec:dyn_timescales}).
Consequently, the trajectory of a cluster on the $(\alpha_\mathrm{global},\Delta \alpha)$ plane serves as a potential diagnostic for constraining its dynamical history and identifying non-standard initial conditions \citep{Webb2016}.

\subsection{Parametrization of evolutionary trends}
\label{sec:parametrize}

To facilitate quantitative comparisons with future studies, we quantify the observed age dependence of $\alpha$ using a smoothly broken linear function over the range $\log(\mathrm{Age/yr}) \in [7, 9.8]$. 
The functional form is given by:
\begin{equation}
\alpha(x) = y_0 + a_1(x - x_0) + \frac{a_2 - a_1}{s} \ln[1 + e^{s(x - x_0)}],
\end{equation}
where $x = \log(\mathrm{Age/yr})$.
This function characterizes the transition from the initial phase (slope $a_1$) to an evaporation-dominated phase (slope $a_2$) at a characteristic turnover age $x_0$, with $s$ controlling the transition sharpness.
We find best-fit parameters $(a_1, a_2, x_0, y_0, s) = (-0.083, 3.7, 8.9, -2.4, 4.9)$ for the global slope $\alpha_\mathrm{global}$, $(0.046, 3.2, 8.7, -2.0, 5.0)$ for $\alpha_{\text{in}}$, and $\allowbreak(-0.18, 3.7, 9.0, -2.6, 8.4)$ for $\alpha_{\text{out}}$. 
These fits quantitatively reproduce the observed trends (Fig.~\ref{fig:mf_fit}) and serve as a reference baseline for future measurements and models.

\subsection{Comparison with N-body simulations}
\label{sec:nbody_valid}

We further compare our observational findings with direct $N$-body simulations using the \textsc{PeTar} code \citep{Wang2020d} (see Appendix \ref{sec:nbody}).
Simulations initialized with a standard Kroupa IMF and no primordial mass segregation reproduce the observed evolutionary tracks of both $\alpha_{\text{global}}$ and $\Delta \alpha$ with high fidelity (Fig.~\ref{fig:mf}).
In contrast, theoretical expectations suggest that significant primordial segregation would persist for several relaxation times before being washed out \citep{Pavlik2019b}; this would yield highly segregated signatures at 10 Myr that are incompatible with our data.
While the oldest clusters ($> 2$ Gyr) deviate from our fiducial simulations ($3500 M_{\odot}$), they are better matched by a more massive model ($7000 M_{\odot}$), consistent with the survival bias, as only initially massive systems have dissolution timescales long enough to survive to these ages (see theoretical estimates in Appendix \ref{sec:dyn_timescales}).

\section{Summary and Discussion}
\label{sec:discussion}

Our study reveals a coherent evolutionary picture for open clusters using a robust census of high-fidelity sample spanning $10^{7}$ to $10^{9.8}$\,yr.
We measure the stellar mass function for $M \ge 0.5 M_\odot$ and its spatial variation (mass segregation) with high precision, utilizing the Bayesian forward-modeling framework MiMO to explicitly account for unresolved binaries and field contamination that often bias traditional methods.

Despite intrinsic cluster-to-cluster variations, the population broadly emerges from the natal gas phase with a canonical Salpeter mass function (mean slope $\alpha_{\text{global}} = -2.29$ with an intrinsic scatter of $0.17$) and an effectively unsegregated structure, defining a baseline initial state for clusters within the local Galactic disk.
The flattened mass functions and strong segregation observed in older clusters are therefore identified as predominantly evolutionary consequences of rapid internal relaxation and delayed external tides ($\sim$600\,Myr), as supported by our direct $N$-body simulations.
Our results provide a robust observational baseline (see \refsec{sec:parametrize} for parameterized fits of $\alpha(t)$ relations) to constrain the physics governing the transition from star-forming clouds to gas-free clusters, and to trace secular dynamics across distinct timescales.

In particular, our findings disfavor star-forming scenarios that predict strong primordial segregation and/or significant IMF variations.
Our observational constraints align with emerging paradigms from recent hydrodynamical simulations.
High-resolution simulations like STARFORGE \citep{Grudic2021,Guszejnov2022b} and TORCH \citep{Polak2025} indicate that, although massive stars often exhibit strong local clustering and rapid segregation within their natal sub-clumps \citep{Allison2009}, this substructure does not necessarily translate into a coherent global segregation signal for typical $\sim$10\,Myr clusters,
contrary to early predictions of preserved segregation \citep{McMillan2007}.
Indeed, our results support the theoretical view that the ``memory'' of any mild primordial inhomogeneities is effectively erased during the chaotic phase of hierarchical assembly of sub-clumps and gas expulsion.
Driven by the resulting rapid potential fluctuations, stars are reshuffled on a global scale by mechanisms such as violent relaxation \citep{Parker2014, Dominguez2017} and dynamical ejection \citep{Polak2025}.
Consequently, clusters emerge into the gas-free regime with a \textit{globally unsegregated} structure.
This natural process effectively resets the dynamical clock, validating the use of non-segregated initial conditions to reconstruct the secular evolution of star clusters.

Lastly, note that the broadly universal initial state reported here is restricted to our sample in the local Galactic disc. Genuine physical variations might be theoretically anticipated \citep{Hennebelle2024} and observationally suggested in extreme environments (e.g., the Galactic Center, \citealt{HosekJr.2018}; massive starburst regions, \citealt{Harayama2008}; and stellar populations with varying metallicity across cosmic time, \citealt{LiJD2023}), which warrants future studies.

\appendix

\section{Illustration of Fitting Formula}
\begin{figure*}[t]
    \centering
    \includegraphics[width=0.85\textwidth]{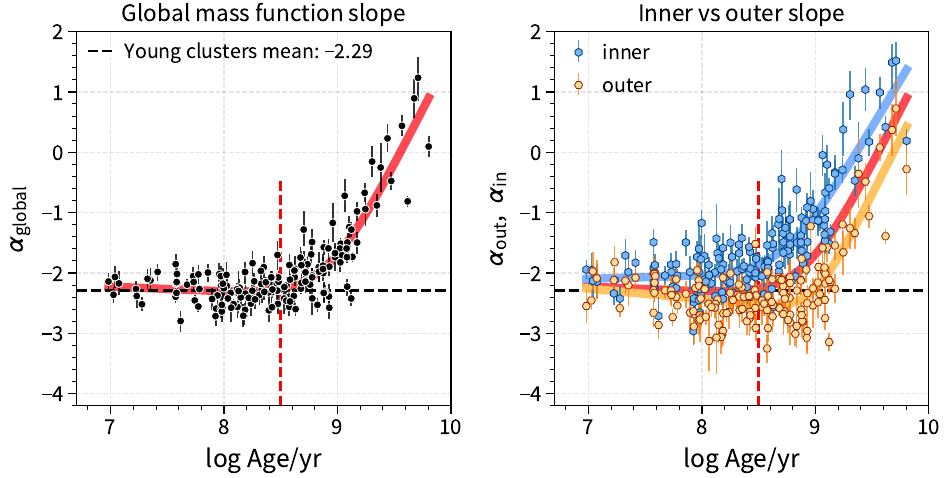}
    \vspace{-0.3em}
    \caption{%
    Same as Fig.~\ref{fig:mf}, but with the fitting formula for mass function slopes from \refsec{sec:parametrize} overlaid. The global mass function fit is also shown in the right panel for reference.
    }
    \vspace{0.5em}
    \label{fig:mf_fit}
\end{figure*}

Fig.~\ref{fig:mf_fit} compares the measured stellar mass function slope with the fitting formula from \refsec{sec:parametrize}.

\section{N-body simulations}
\label{sec:nbody}

To validate the dynamical origin of the observed trends, we performed direct $N$-body simulations using the code \textsc{PeTar} \citep{Wang2020d}, which incorporates stellar evolution, binary interactions, and mergers. 
These simulations serve as fiducial benchmarks; a comprehensive exploration of the vast parameter space is beyond the scope of the present work.

In the fiducial run, the cluster is initialized with a total mass of $M\sim 3500\,M_\odot$ ($N\sim 2000$ stars) within the tidal radius, following a Kroupa \citep{Kroupa2001} IMF ($0.08$--$90\,M_\odot$; $\alpha = -2.3$ for $M > 0.5\,M_\odot$).
Stars are distributed according to a Plummer model with a half-mass radius of $r_{\text{hm}}=1$\,pc. 
The cluster is placed in a Milky Way potential on a circular orbit of 8\,kpc.
We assume no primordial mass segregation and binaries to isolate the effects of dynamical evolution.

To cover a wider dynamical range, we performed ancillary runs varying the half-mass radius ($0.5$ and $3$\,pc), 
orbital radius ($6$ and $10$\,kpc), and total mass ($7000\,M_\odot$), changing only one parameter at a time.
Systems were evolved for 3\,Gyr (6\,Gyr for the massive run) or until the number of member stars fell below 100. 

We analyzed snapshots at 1\,Myr intervals, calculating the MF slopes for stars with $M > 0.5\,M_\odot$ (excluding merger products and compact remnants) within the instantaneous tidal radius.
Measurements were performed both globally and separately inside/outside the projected half-number radius, yielding the synthetic $\alpha_{\text{global}}$, $\alpha_{\text{in}}$, and $\alpha_{\text{out}}$ trajectories.
The simulations demonstrate agreement with the observed evolutionary trends, reproducing both the delayed onset of global MF flattening and the rapid establishment of mass segregation.

\section{Theoretical dynamical timescales}
\label{sec:dyn_timescales}

We estimate characteristic dynamical timescales to provide a physical reference. 
A fiducial dynamical (crossing) timescale, $t_{\mathrm{dyn}}$, is 
\begin{equation}
  t_{\mathrm{dyn}} = \sqrt{\frac{r_{\mathrm{h}}^3}{GM}}
\end{equation}
where $r_{\mathrm{h}}$ is the half-mass radius and $M$ is the total cluster mass. 
The half-mass relaxation time, $t_{\mathrm{rh}}$, which governs the rate of energy exchange and memory loss of initial conditions, follows \citep{Spitzer1987}
\begin{equation}
  t_{\mathrm{rh}} \approx \frac{0.138 N}{\ln (\gamma N)} t_{\mathrm{dyn}}
\end{equation}
where $N$ is the number of stars and $\gamma \simeq 0.02$ for multi-mass clusters \citep{Giersz1996}. 
Massive stars sink to the center on the mass segregation timescale, $t_{\mathrm{seg}}$.
For stars of mass $m_{\mathrm{target}}$ in a cluster with mean stellar mass $\bar{m}$, this is approximated by \citep{Spitzer1969}
\begin{equation}
  t_{\mathrm{seg}} (m_{\mathrm{target}}) \approx \frac{\bar{m}}{m_{\mathrm{max}}} t_{\mathrm{rh}}.
\end{equation}
Clusters preferentially lose low-mass stars (which acquire higher velocities) via internal evaporation ($t_{\mathrm{evap}} \sim 100 \, t_{\mathrm{rh}}$ in isolation) and environmental tidal stripping.
In the Galactic tidal field, the dissolution timescale, $t_{\mathrm{diss}}$ (time required for 95\% mass loss), follows the relation \citep{Baumgardt2003}
\begin{equation}
\begin{split}
  t_{\mathrm{diss}} & \propto t_{\mathrm{rh}}^x t_{\mathrm{dyn}}^{1-x} \\
  & \approx \beta \left[ \frac{N}{\ln (\gamma N)} \right]^x 
  \frac{R_G / \mathrm{kpc}}{V_G / 220 \, \mathrm{km \, s}^{- 1}} (1 - \varepsilon)
  \, \mathrm{Myr},
\end{split}
\end{equation}
where the coefficient $\beta \simeq 1.91$ and the tidal scaling exponent $x = 0.75$ for typical open clusters orbiting at Galactocentric distance $R_G$ with circular velocity $V_G$ and orbital circularity $\varepsilon$. 

Although $t_\mathrm{seg}$ and $t_\mathrm{diss}$ are correlated, they exhibit distinct dependencies on internal structure and the external tidal field.
This differential dependence breaks the degeneracy between internal and external dynamical drivers, making the joint analysis of $\alpha_\mathrm{global}$ (evaporation) and $\Delta \alpha$ (segregation) a promising diagnostic for constraining cluster evolution \citep{Webb2016}.
In this work, we utilize these timescales to provide a physical context for the observed trends, leaving detailed modeling of individual evolutionary histories for future studies.

For our fiducial simulation ($M_{\mathrm{ini}} = 3500 \, M_{\odot}$ at $R_G = 8$\,kpc), early stellar evolution reduces the system to $M \simeq 3000 \, M_{\odot}$ and $N \simeq 2000$ by 5\,Myr. 
With $r_{\mathrm{h}} = 1$\,pc, we derive characteristic timescales of $t_{\mathrm{dyn}} \approx 0.3$\,Myr, $t_{\mathrm{rh}} \approx 20$\,Myr, and $t_{\mathrm{seg}} \approx 10$\,Myr (for $3 \, M_{\odot}$ stars), with a dissolution time of $t_{\mathrm{diss}} \approx 1700$\,Myr. 
Significant global MF evolution begins at ${\sim}30\%$ mass loss \citep{Vesperini2009}, corresponding to a timescale of $0.3 t_{\mathrm{diss}} \approx 520$\,Myr (assuming linear mass loss; \citealt{Baumgardt2003,Lamers2013}).
Similarly, for the more massive cluster ($M_{\mathrm{ini}} = 7000 \, M_{\odot}$), we obtain $(t_{\mathrm{dyn}}, t_{\mathrm{rh}}, t_{\mathrm{seg}}, t_{\mathrm{diss}},0.3 t_{\mathrm{diss}}) \approx (0.2, 24, 12, 2500, 760)$\,Myr.
Although specific timescales vary with cluster parameters, these theoretical estimates show general agreement with our observational and simulated results.

\begin{acknowledgments}
This work is supported by the National Natural Science Foundation of China (NSFC 12303026 and 12273091); the Science and Technology Commission of Shanghai Municipality (22dz1202400); the Young Data Scientist Project of the National Astronomical Data Center, the Program of Shanghai Academic/Technology Research Leader and the China Manned Space Program through its Space Application System. 
ZZ.L. acknowledges the Fundamental Research Funds for the Central Universities, Nanjing University (KG202502), and the Marie Skłodowska-Curie Actions Fellowship (101109759, CuspCore) under Horizon Europe.
L.W. thanks the support from the National Natural Science Foundation of China through grant 21BAA00619 and 12233013, the High-level Youth Talent Project (Provincial Financial Allocation) through the grant 2023HYSPT0706, the one-hundred-talent project of Sun Yat-sen University, the Fundamental Research Funds for the Central Universities, Sun Yat-sen University (22hytd09).
\end{acknowledgments}

\begin{contribution}
L.L. led the project and data analysis and drafted the manuscript. Z.Z.L. led the theory, simulation design, visualization, and manuscript revision. Z.Y.S. provided conceptual and technical input. Z.P.Z. and L.W. performed and analyzed the $N$-body simulations.
\end{contribution}

\section*{Data and Code Availability}
The full MiMO catalog and the ``MF primer'' sample (including derived parameters, full likelihood chains, and the model isochrone files) are available on the China-VO PaperData service of the National Astronomical Data Center (NADC), China \citep{MiMOcat}.
The MiMO source code is available from the same repository, with a copy on GitHub (\url{https://github.com/luly42/mimo}).
Other materials supporting this study, including code and simulation evolution tracks to reproduce Figs.~\ref{fig:mf} and \ref{fig:mf_fit}, are available from the corresponding authors upon request.

\bibliographystyle{aasjournalv7}

\bibliography{main}{}

\end{CJK*}
\end{document}